\documentclass[11pt,draftclsnofoot, onecolumn, peereview]{IEEEtran}
\usepackage{graphicx,cite,amssymb,amsmath,stmaryrd,marvosym,textcomp}
\renewcommand{\QED}{\QEDopen}

\usepackage{multirow}
\usepackage{booktabs}
\usepackage{bigstrut}
\usepackage{graphicx}
\usepackage{subfigure}
\usepackage{array}
\usepackage{color}

\usepackage{url}

\begin{document}

\title{\mbox{}\\
Artificial Intelligence Enabled Wireless Networking for 5G and Beyond: Recent Advances and Future Challenges}

\author{Cheng-Xiang~Wang,~\IEEEmembership{Fellow,~IEEE},
Marco Di Renzo,~\IEEEmembership{Senior Member,~IEEE},
Slawomir Sta\'nczak,~\IEEEmembership{Senior Member,~IEEE},
Sen Wang, and Erik G. Larsson,~\IEEEmembership{Fellow,~IEEE}
\thanks{C.-X. Wang (corresponding author) is with the National Mobile Communications Research Laboratory, School of Information Science and Engineering, Southeast University, Nanjing, 210096, China, and also with the Purple Mountain Laboratories, Nanjing, 211111 (e-mail:chxwang@seu.edu.cn).}
\thanks{M. D. Renzo is with Université Paris-Saclay, CNRS, CentraleSupélec, Laboratoire des Signaux et Systèmes, 91190, Gif-sur-Yvette, France (e-mail: marco.direnzo@l2s.centralesupelec.fr).}
\thanks{S. Sta\'nczak is with the Fraunhofer Institute for
Telecommunications, Heinrich Hertz Institute, 10587 Berlin, Germany, and also with the Technical University of Berlin, 10623 Berlin, Germany (e-mail: slawomir.stanczak@hhi.fraunhofer.de).}
\thanks{S. Wang is with the Institute of Sensors, Signals and Systems, School of Engineering \& Physical Sciences, Heriot-Watt University, Edinburgh, EH14 4AS, U.K. (e-mail: s.wang@hw.ac.uk).}
\thanks{E. G. Larsson is with the Department of Electrical Engineering (ISY), Link\"oping University, SE-581 83 Link\"oping,  Sweden (e-mail: erik.g.larsson@liu.se).}
\thanks{This work was supported by the National Key R\&D Program of China under grant 2018YFB1801101, the National Natural Science Foundation of China (NSFC) under grant 61960206006, the Fundamental Research Funds for the Central Universities under grant 2242019R30001, and the EU H2020 RISE TESTBED project under grant 734325. The authors would also like to thank Dr. Yu Fu from Heriot-Watt University and Dr. Jie Huang from Southeast University for their valuable comments on this article.}}

\maketitle
\vspace{-2.0cm}
\begin{abstract}
The fifth generation (5G) wireless communication networks are currently being deployed, and beyond 5G (B5G) networks are expected to be developed over the next decade. Artificial intelligence (AI) technologies and, in particular, machine learning (ML) have the potential to efficiently solve the unstructured and seemingly intractable problems by involving large amounts of data that need to be dealt with in B5G. This article studies how AI and ML can be leveraged for the design and operation of B5G networks. We first provide a comprehensive survey of recent advances and future challenges that result from bringing AI/ML technologies into B5G wireless networks. Our survey touches different aspects of wireless network design and optimization, including channel measurements, modeling, and estimation, physical-layer research, and network management and optimization. Then, ML algorithms and applications to B5G networks are reviewed, followed by an overview of standard developments of applying AI/ML algorithms to B5G networks. We conclude this study by the future challenges on applying AI/ML to B5G networks.
\end{abstract}

\date{\today}
\renewcommand{\baselinestretch}{1.2}
\thispagestyle{empty} \maketitle \thispagestyle{empty}
\newpage
\setcounter{page}{1}

\IEEEpeerreviewmaketitle

\section{Introduction}
Global demand for wireless communication networks continues to increase, mainly due to the ever-growing numbers of wireless users and new emerging wireless services. The fifth generation (5G) and beyond 5G (B5G) wireless networks are expected to be developed in the future and offer higher data rates, improved coverage, better cost efficiency, resource utilization, security, adaptability, and scalability \cite{Wang14}. Artificial intelligence (AI) technologies have the potential to efficiently solve unstructured and seemingly intractable problems involving large amounts of data that need to be dealt with in the design and optimization of 5G and B5G wireless networks.

AI is ``the simulation of human intelligence processes by machines, especially computer systems" \cite{Bi15}. It is usually defined as the science of making computers perform tasks that require intelligence like humans. Whereas AI is a broader concept of machines being able to carry out tasks smartly, machine learning (ML) is a current (probably the most popular) application of AI that enables machines to learn from large amounts of data and act accordingly without being explicitly programmed. As a special type of ML, deep learning studies artificial neural networks (ANNs) that contain more than one hidden layer to ``simulate" the human brain. Currently, deep learning is one of the most widespread ML methods as it has successfully been applied to different fields such as the computer vision, speech recognition, and bioinformatics.

AI technologies will not only reduce or even replace manual efforts for the network development, configuration, and management, but also deliver better system performance, reliability, and adaptability of communication networks by making real-time robust decisions based on predictions of the networks and users' behavior. ML, as a typical AI technologies, is widely expected to rapidly become a key component of B5G communication networks. It will make a full use of the big data to overcome the challenges of designing and operating B5G networks. Potential benefits of introducing ML into communication systems include the following. Firstly, channel and interference models are extremely complicated in reality due to the dynamic nature of wireless communication channels, especially in B5G scenarios. ML techniques may automatically extract the unknown channel information by learning from the communication data and prior knowledge. Secondly, as the density of wireless access points continues increasing, there is an urgent need for global optimization of communication resources and fine tuning of system settings. However, the enormous amount of resources, system parameters to be optimized, and their coupled correlations render these tasks notoriously difficult to solve using existing approaches. In contrast, sophisticated ML algorithms, e.g., deep learning and probabilistic learning methods, may be able to model the highly non-linear correlations and estimate (sub-)optimal system parameters. Lastly, ML will realize learning-based adaptive configuration of networks by finding out behavioral patterns and responding timely and flexibly to various scenarios, e.g., anticipating traffic and planning ahead rather than simply reacting to unexpected events.

Current cellular networks designed and operated based on previous postulates may systematically fail to enable future communication services, since they cannot keep pace with the data explosion and the underlying complexity of the generated data while guaranteeing the required capacity, reliability, and adaptability. Thus, the network cannot quickly react and anticipate events that might deteriorate communication services in real-time. However, as most AI algorithms and applications are not specifically designed for wireless communication networks, it is hard to directly apply existing AI algorithms to B5G networks. 

\begin{figure}[p]
\begin{center}
\vspace{1.5cm}
 \includegraphics[width=6in]{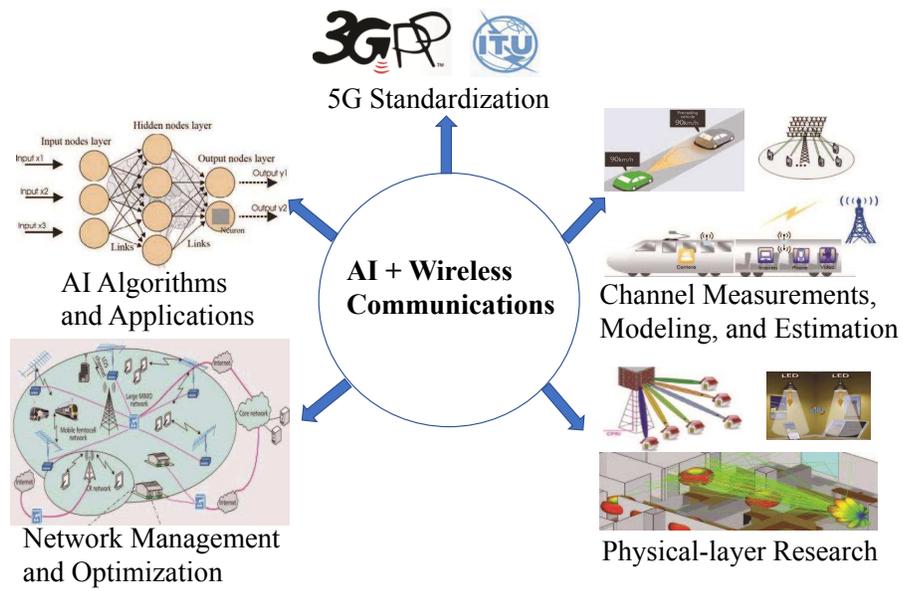}\\
  \caption{Research aspects that bring AI technologies into B5G wireless networks.}\label{5G AI}
\end{center}
\end{figure}

Compared to earlier survey papers such as \cite{Bi15}, this article aims to explore the advantages of combining AI technologies with B5G wireless networks, leveraging the potential of AI technologies to tackle challenges that cannot be efficiently addressed using conventional communication technologies. This article focuses on the following five aspects that bring AI technologies into B5G wireless networks as shown in Fig. \ref{5G AI}. Accordingly, the remainder of the paper is organized as follows. In Section \ref{Sec2}, we discuss channel measurements, modeling, and estimation for B5G networks using AI technologies. Section \ref{Sec3} studies physical-layer researches for B5G networks using AI technologies. In Section \ref{Sec4}, we provide a survey of network management and optimization for B5G networks using AI technologies. AI algorithms and applications to B5G networks are given in Section \ref{Sec5}. Section \ref{Sec6} provides an overview of standard developments of AI technologies and ML algorithms for B5G networks. Conclusions are drawn and future challenges are discussed in Section~\ref{Sec7}.

\section{Channel Measurements, Modeling, and Estimation for B5G Networks Using AI Technologies}\label{Sec2}

\subsection{Channel Measurement Data Processing and Channel Modeling}
For B5G wireless communication systems, the diversity of frequency bands, including sub-6~GHz, millimeter wave (mmWave), terahertz (THz), and optical bands, has made channel modeling more complex. To address the B5G channel modeling requirements, the existing channel models are extended with a much higher computational complexity. When modeling for new scenarios, channel measurements must be conducted to understand new channel characteristics, which is a time-consuming task. Apart from a recent work in \cite{Ald19}, there are very few investigations that studied the benefits of applying AI to channel modeling and most existing works only employ very simple AI techniques on a very limited part of the channel modeling process. There is no work that comprehensively investigates the application of AI technologies to channel measurements.

Due to the high complexity to model the signal propagation in diverse scenarios, conventional methods make many assumptions and approximations to simplify the processing and modeling methods. Wireless channel features can be extracted from the huge amount of existing measurement data and, at the same time, the channel modeling problem can be tackled in a data-driven manner, seamlessly integrating with model-based methods. A good balance of the accuracy-complexity trade-off of both processing and modeling techniques will be maintained.

\begin{figure}[p]
\centering
\begin{minipage}[t]{0.48\linewidth}
\centerline{\includegraphics[width=3in]{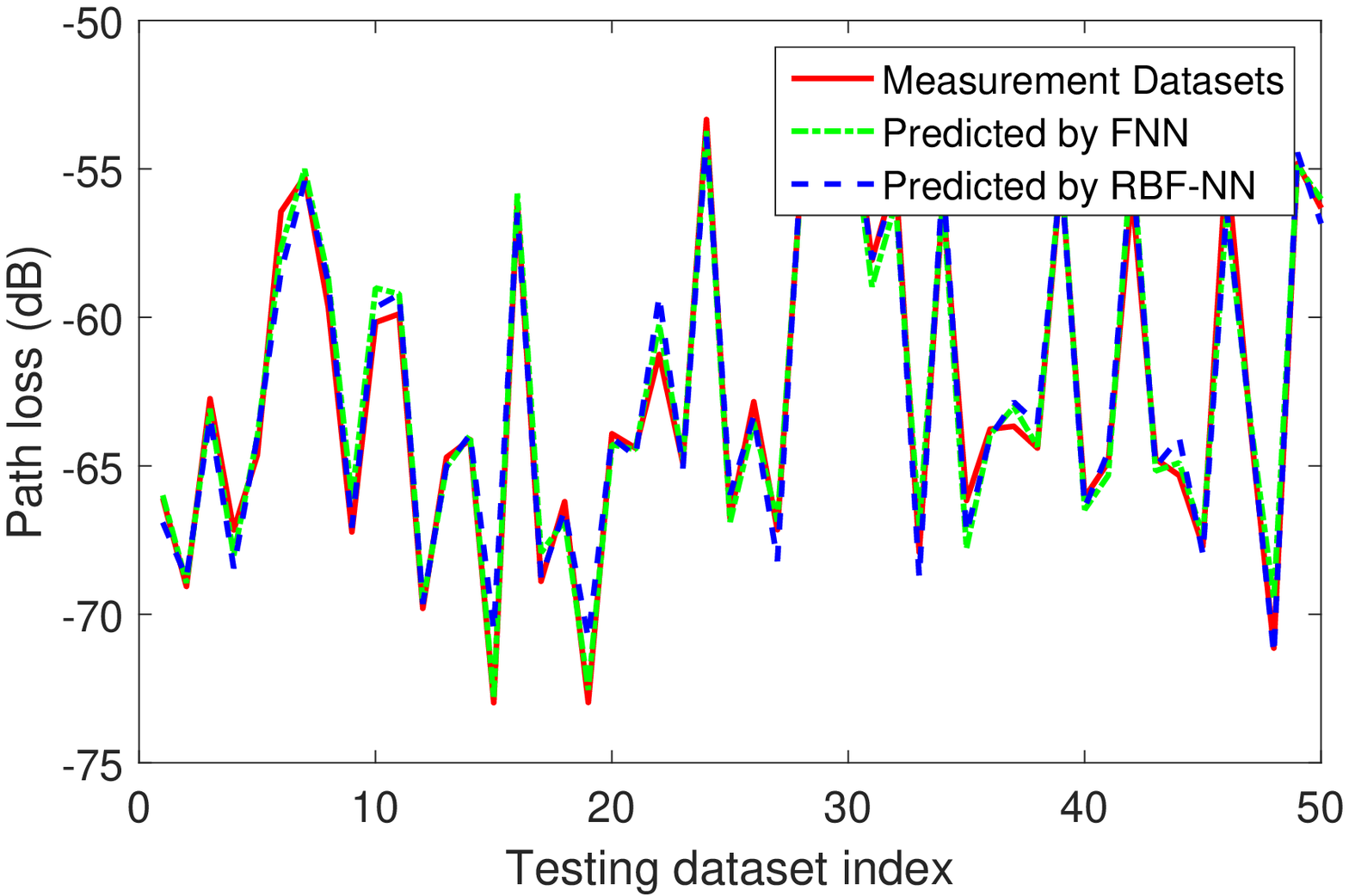}}
\footnotesize \centerline{(a) Measured and predicted path loss}
\end{minipage}
\begin{minipage}[t]{0.48\linewidth}
\centerline{\includegraphics[width=3in]{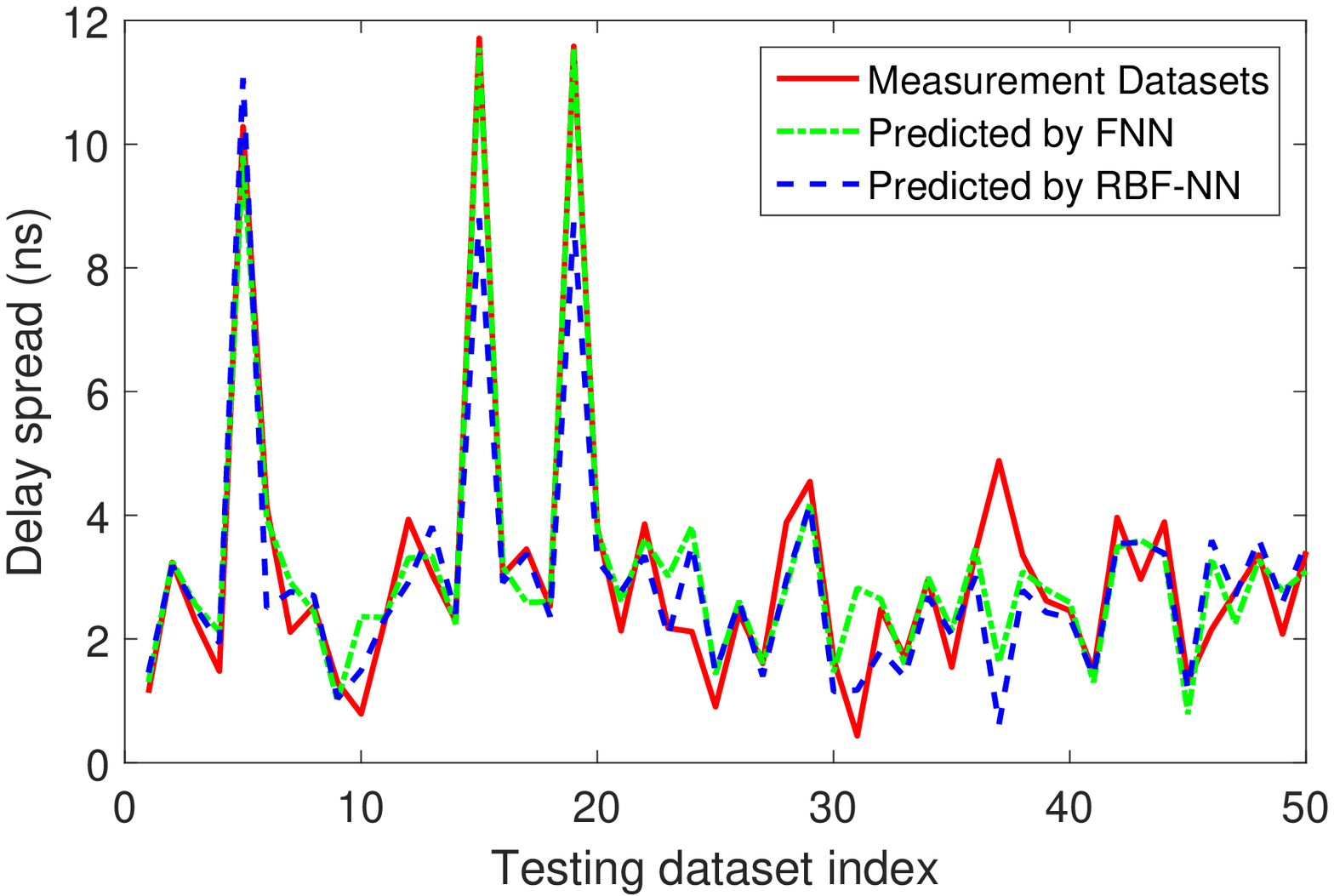}}
\footnotesize \centerline{B Measured and predicted RMS DS}
\end{minipage}
\caption{(a) Measured and predicted path loss, and B measured and predicted RMS DS.}
\label{fig:Fig_1}
\end{figure}

ML can be utilized to channel features prediction, channel impulse response (CIR) modeling, multipath component (MPC) clustering, channel parameter estimation, and scenario classification, based on channel measurement data and environment information. Authors in \cite{Huang18} proposed a big data enabled channel model based on both feed-forward neural network (FNN) and radial basis function neural network (RBF-NN). It can predict channel statistical properties including the received power, root mean square (RMS) delay spread (DS), and RMS angle spreads (ASs) with input parameters of transmitter (Tx) and receiver (Rx) coordinates, Tx--Rx distance, and carrier frequency. The performances of FNN and RBF-NN were fully compared based on both real channel measurement data and synthetic data. An example of the measured and predicted path loss and RMS DS is shown in Fig. \ref{fig:Fig_1}. Both FNN and RBF-NN show good potential for channel modelling. In \cite{Ma17}, the ANN was applied to remove the noise from measured CIR, and the principal component analysis (PCA) was utilized to exploit the features and structures of the channel and model the CIR. In \cite{He18}, several clustering algorithms were investigated for MPC clustering and tracking, including K-means, fuzzy C-means (FCM), and density-based spatial clustering of applications with noise (DBSCAN). In \cite{Li17}, the convolutional neural network (CNN) was used to automatically identify different wireless channels and help decide which relevant wireless channel features should be used. The MPC parameters like amplitude, delay, and Doppler frequency were extracted and used as input parameters in the CNN, and the output of the CNN was the class of the wireless channels.

\subsection{Channel Estimation Associated with ML}
In wireless communications, the channel state information (CSI) can be acquired through blind and pilot-based channel estimation techniques. However, blind channel estimation extracts statistical properties by using abundant received symbols. For the pilot-based technique, with the deployment of 5G key technologies, pilot overhead, non-linear channel, and high mobility channel, etc., are challenges to be conquered in channel estimation. For example, the pilot overhead can be intolerable for massive multiple-input multiple-output (MIMO) channels and ultra-dense networks (UDNs). Thus, the trade-off between pilot length and channel estimation accuracy should be considered. The non-linear characteristics of visible light communication (VLC) channel and mmWave channel make it hard to get accurate CSI. In addition, with the development of high speed railways (HSRs), accurate channel estimation is important to guarantee the quality-of-service (QoS) and efficient information transmission.

Researchers resort to ML techniques to solve above mentioned problems. To address the channel estimation in a fast fading time-varying multipath channel, a two dimensional (2D) non-linear complex support vector regression (SVR) based on a RBF kernel was proposed to achieve accurate channel estimate \cite{Channelestimation4}. In \cite{He18_2}, a deep learning based channel estimation algorithm was proposed for beamspace mmWave massive MIMO systems. It can learn
channel structure and estimate channel from a large number of training data. In \cite{Tang18}, an off-grid sparse Bayesian learning based channel estimation algorithm was proposed for mmWave massive MIMO uplink. It can identify the angles and gains of the scatterer paths by exploiting spatial sparse structure in mmWave channels. 

For this topic, one future direction is the generalized ML-based channel estimation scheme, which can be directly used in different scenarios without further training. In order to build this generalized scheme, vast amount of pre-collected communication data have to be used by machine/deep learning algorithms to learn the channel feature of different environments.

\section{Physical-Layer Research for B5G Networks Using AI Technologies}\label{Sec3}

\subsection{Large-scale Sensing via Massive Radio Interfaces}
The use of large antenna arrays offer not only the unprecedented performance in terms of reliable and high-rate communications (as exploited in massive MIMO), but also provide enormous amounts of baseband-level data that can be used to make inferences about the environment. Emerging more novel use cases include inference problems, for example, detection of the presence of moving objects, estimation of the amount of traffic on a road, counting of the number of persons in a room, or guarding against intrusion in protected spaces. Particular technical challenges that lie within reach are the sensing of open spaces, indoor venues, and even through-the-wall. There are emerging commercial use cases and also many applications in security, surveillance, and monitoring.

ML algorithms are particularly suitable to analyze the vast amounts of data generated by large antenna arrays, especially massive MIMO arrays, as typically parametric models are unavailable or inaccurate, hence classical estimation/detection algorithms are inapplicable. More specifically, in terms of algorithmic approach, deep learning networks and methodology from image processing and video analytics may offer the most promising path. It is important to note the distinction to conventional radar imaging, where the objective is to create an image or map of the environment, whereas the goal of emerging large-scale sensing is to extract specific features of the dynamics of the environment and make inference about specific phenomena \cite{Li18}.

Important future research directions should include both pertinent physical modeling work and the construction of an algorithmic foundation that exploits relevant ML tools. Trained deep neural networks represent an important technology component in this regard, but also various forms of dictionary learning
might be used. Simulated channel models should be used for evaluation, along with experimentally obtained real data. Through the use of these techniques, research along this direction could significantly advance the state-of-the-art in sensing of open spaces, indoor venues, and through-the-wall and accomplish inference tasks that are impossible with conventional model-based signal processing. Another application that may benefit from the technology is gesture recognition, especially when implementing sensing at higher frequencies.

\subsection{Signal Processing}
Massive MIMO technologies have been adopted in 5G communication systems. It is one of the obvious use cases that AI can be deployed. Although massive MIMO has many advantages, such as spectrum efficiency, energy efficiency, security, and robustness, it can produce a large volume of data. For example, in channel measurements, a massive MIMO system with 32$\times$56 antennas and 100 MHz bandwidth can produce data larger than 32 Gbyte. Both detection and channel estimation for massive MIMO systems are usually time-consuming processes and require great computational power. The big data property of massive MIMO system makes researchers think of ML methods. In \cite{Sure2016}, the large amount of data generated from massive MIMO system is represented by large random matrices and analyzed using the single ring law. One of the challenges for massive MIMO system is pilot contamination, which can make a significant impact on the performance of massive MIMO systems. The pilot contamination stems from the pilot interference between adjacent cells and can limit the ability of systems to obtain accurate CSI. As the number of antennas increases, channels in beamspace are approximately sparse, i.e., most of the MPC power results from a few paths gathered into clusters in the space and the channel matrix contains a small number of nonzero elements \cite{Gao18}. Based on the sparsity property of channels in beamspace, authors in \cite{Wen2015} obtain the CSI of massive MIMO systems using the sparse Bayesian learning method. Compared with the conventional CSI estimators, the Bayesian learning method can achieve a better performance in terms of pilot contamination. The sparse recovery problem is an important research issue for the Bayesian compressive
sensing. It aims to estimate a non-negative compressible vector from a set of noiseless measurements. 

\subsection{Data-driven Localization in Wireless Networks}
Accurate positioning is valuable for context awareness and location based network management and services. Most of current wireless positioning techniques use channel information and fingerprinting to estimate locations. In reality, the channel information needs to be frequently updated to reflect the true channel characteristics since they are susceptible to a variety of dynamic and time-varying transmission impediments, e.g., path loss, interference, and blockage. This periodic and long-term maintenance is time-consuming and labor-intensive, especially for large-scale B5G systems. As the amount and diversity of sensing and communication data dramatically increase in the B5G systems, data-driven localization is a promising solution, i.e., positioning devices and users through learning from raw sensing and communication data with ML algorithms. The data-driven localization algorithms will not only be self-adaptive to the real-time dynamic transmission impediments, but also evolve over time by consistently learning from data. The wireless channel locations can be continuously updated and improved by automatically learning from the crowd-sourced big data from a vast number of mobile devices. Benefiting from the accurate localization results, users will enjoy better location based services in return.

\section{Network Management and Optimization for B5G Networks Using AI technologies}\label{Sec4}

\subsection{From Model-based to Data-driven Optimization of UDNs}
The current approach to the management and optimization of cellular networks is based on ``models". This approach is used across all the network functionalities, from the design of the physical-layer to the deployment of network infrastructure. However, it is insufficient for the design of future networks, which will be based on multiple and diverse radio access technologies, ultra-densely deployed, and have to serve a broad class of applications and requirements. A so complicated network eco-system cannot be optimally designed and orchestrated based on ``models" that reproduce only in part actual network deployments and that are not accurate in practice. For example, let us consider a typical on-demand network deployment that relies on the use of unmanned aerial vehicles (UAVs) for rescue operation, disaster recovery, etc. Such a network needs to be deployed in an ad hoc manner and cannot rely on models that do not even exist for the specific case of interest. Based on these premises, there is a compelling need for radically changing the way future networks will be engineered and optimized. The complexity of future networks and the broad set of requirements that they need to fulfill necessitate them to go beyond the concept of models for network design and to exploit the large availability of ``data". In this context, a paradigm-shift for the efficient design of B5G networks is necessary: To leverage AI and ML in order to take advantage of big data analytics to enhance the situational awareness and overall network operation of future networks. AI can parse through massive amounts of data generated from multiple sources such as wireless channel measurements, sensor readings, and drones and surveillance images, to create a comprehensive operational map of the massive number of devices within the network. It can be exploited to optimize various functions, such as fault monitoring and user tracking, across the wireless network. Resource management mechanisms based on AI will be able to operate in a fully online manner by learning the states of the wireless environment and the network's users in real time. Such mechanisms will be able to continuously improve their own performance over time which in turn will enable more intelligent and dynamic network decision making.

\begin{figure}[p]
\centering
\includegraphics[width=4.5in]{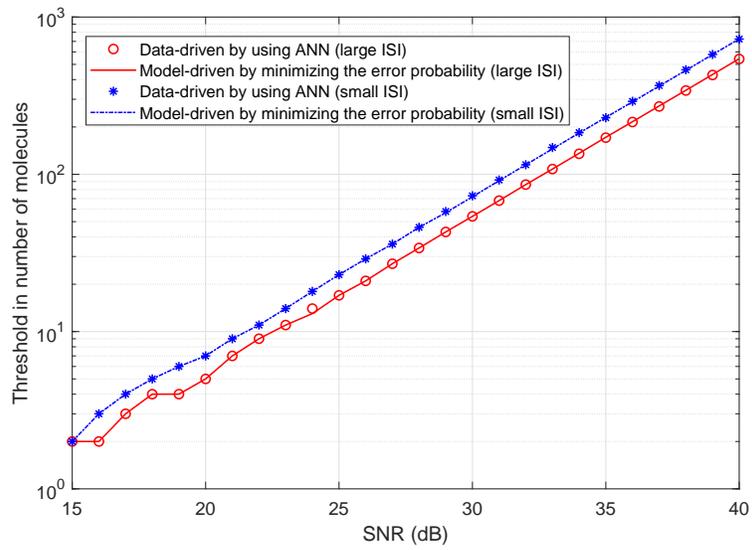}
\caption{Optimal demodulation threshold for small and large values of the ISI.}
\label{fig:Fig_3}
\end{figure}

In order to substantiate the potential of using ANNs for the design of communication networks, we consider the optimization of a typical threshold-based demodulator for Poisson channels, which find applications in optical and molecular communication networks that are known to be difficult to model in the presence of inter-symbol interference (ISI). We compare the typical approach employed in communications, where the system/channel model is assumed to be perfectly known and the optimal demodulation threshold is obtained by minimizing the analytical expression of the error probability that accounts for the ISI, against a data-driven approach, where nothing is known about the channel model and ANNs are used in order to learn the best demodulator and thus, the optimal demodulation threshold without prior information on the system model. As far as the data driven-approach is concerned, the ANN is trained by using supervised learning. In particular, we use the Bayesian regularization back propagation method, which updates the weights and biases of the ANN by using the Levenberg-Marquardt optimization algorithm. Fig. \ref{fig:Fig_3} shows the optimal demodulation threshold for two values of the ISI (small and large values) in different signal-to-noise ratios (SNRs). In both cases, we note that a data-driven approach provides one with the same demodulation threshold and consequently the same error probability, by dispensing the system designer to perfectly know the system/channel model.

\subsection{Proactive Wireless Networking for Online Software Networks Orchestration}
Recently, the industry has witnessed the increasing maturity of software-defined networking (SDN) and network function virtualization (NFV), which constitute fundamental enabling technologies to realize the 5G PPP vision of software/programmable networking. With the evolution of SDN and NFV, 5G cellular networks have advocated a revolutionary concept called network slicing (NS). Instead of building dedicated networks for different services, NS allows operators to intelligently create customized network pipes to provide optimized solutions for different services that require diverse functionalities, performance metrics, and isolation criteria. Specifically, mobile edge computing, edge caching, etc., have potentially evolved to replace its forwarding-only functionality to an area equipped with storage, memories, and computational power capabilities. Enabling future cellular architectures with NS is a fundamental necessity for optimal network orchestration and for offering services with so diverse requirements, notably enhanced mobile broadband (eMBB) that needs bandwidth-consuming and throughput-driving to new services such as ultra-reliable and low-latency communications (uRLLC) and massive machine-type communications (mMTC). Today’s networks and even 5G networks are conceived, designed, and optimized based on the reaction principle which passively responds to incoming demands and serves them when requested. This principle is not adequate for the new service capabilities that future networks need to provide. The future networks will be heterogeneous software-defined networks. Even different services are logically independently operated, their data traffic will finally be mixed, which can result in a highly-dynamic and unmanageable manner to the reaction principle. Besides, for future networks, the requirements of some services cannot be fulfilled by the current reaction principle. For example, uRLLC applications may not accept the delay associated with this reaction principle. On the other hand, future networks need prediction capabilities, which enable them to anticipate the future and proactively allocate network resources. In a proactive approach, rather than passively responding to incoming demands and serving them when requested, network architectures with NS can predict traffic patterns and determine future off-peak times on different spectrum bands so that incoming traffic demands can be properly allocated over a given time window. Predicting the users' behaviors will result in a better utilization of the network resources and will allow us to optimally allocate end-to-end network slices in an online fashion. This paradigm-shift from reactive to proactive network design can only be made possible with the aid of AI and ML techniques.

\section{AI Algorithms and Applications for B5G Networks}\label{Sec5}

\subsection{Distributed ML Algorithms for B5G Networks}
\begin{figure}[p]
\centering
\includegraphics[width=4.5in]{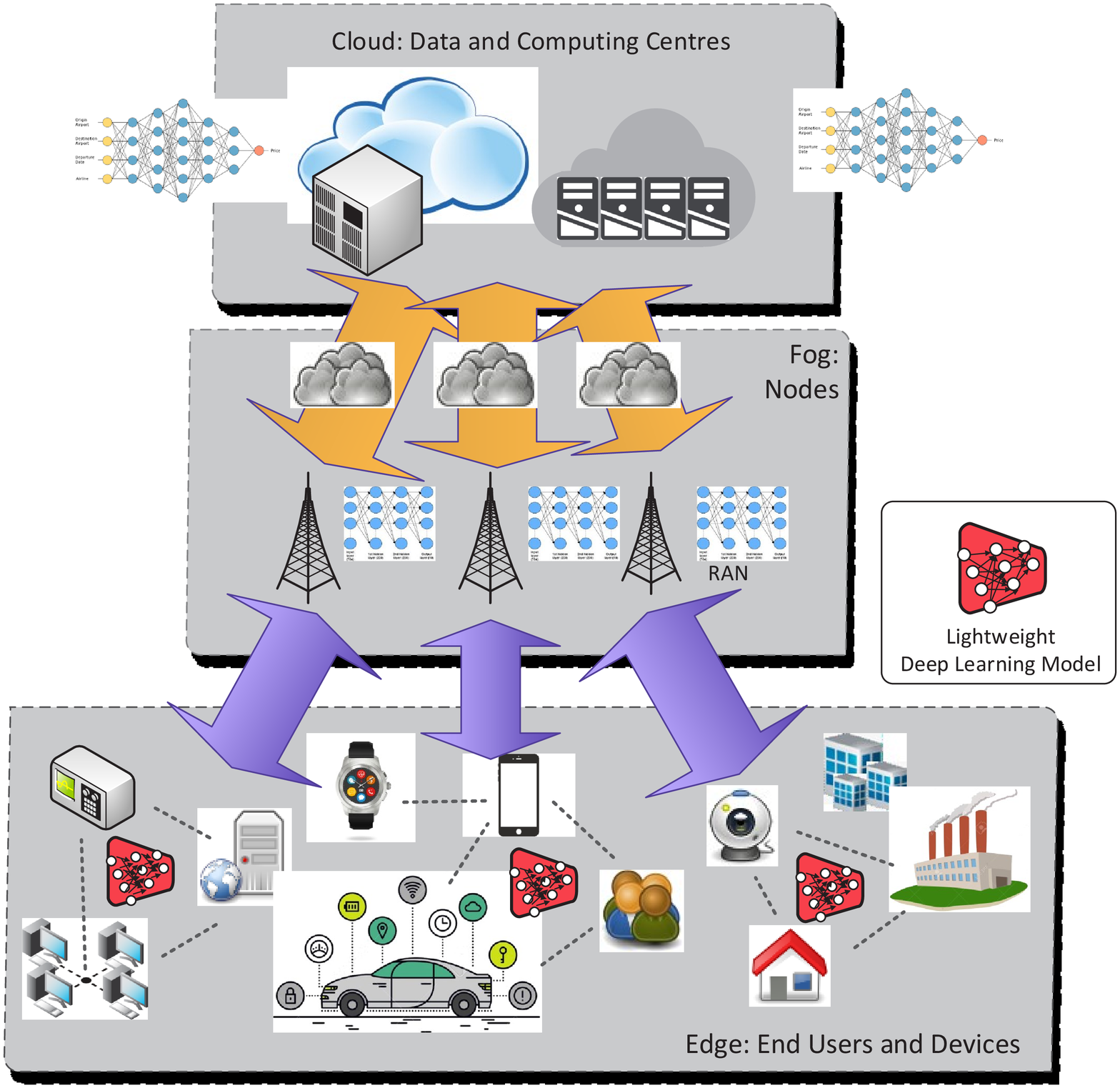}
\caption{Application of deep learning in cloud, fog, and edge computing networks.}
\label{fig:Fig_4}
\end{figure}

In current communications applications, signal processing and ML algorithms are typically executed centrally. An archetype architecture is the cloud-radio access network (C-RAN), where joint estimation and data processing for all network devices is performed at a central unit (e.g., the cloud). In the presence of a large number of devices and communication limitations on the fronthaul/backhaul links, however, various network functions should be executed locally or with minimal information exchange with the cloud. Therefore, of central importance is the provisioning of a decentralized functional architecture, which adapts dynamically on the network requirements \cite{Fu18}. As shown in Fig. \ref{fig:Fig_4}, lightweight deep learning model can be applied to cloud, fog, and edge computing networks. The cloud network is the data and computing center, the fog network includes many nodes, and the edge network contains enormous end users and devices. In parallel, there is the need for decentralized learning, classification, and signal processing algorithms, which seamlessly adapt to the number and the type of the information sources, considering the available communication bandwidth. In the presence of a dynamic edge computing architecture, the advantages of decentralized and centralized algorithms should be combined, thereby trading-off complexity, latency, and reliability. This requires integration and further development of methods for data fusion, compression, and distributed decision-making. 

In the distributed setting, there is also the need of developing solutions that are capable of learning the relationships between the network entities and their time evolution. Since dynamical network inference is a complex task in general, scalable solutions are required. Therefore, it is necessary to evaluate the potential of online learning methods, such as kernel-based adaptive filters, high-dimensional set-theoretic algorithms, and other robust statistical estimation methods. Additional examples include Bayesian approaches in conjuncture with approximate inference methods, such as approximate message passing and generalizations therein.

\subsection{ML Algorithms for Ultra-fast Training and Inference}
ML algorithms are mostly designed for systems and applications which do not need to achieve high-frequency performance. Unfortunately, this is not the case in the context of B5G networks, which must ensure high data processing rates for ultra-low latency. This imposes a strict requirement on the speeds of training and, particularly, inference of ML models. Therefore, one big challenge is to develop ML algorithms with ultra-fast training and inference capability for future wireless communications.

There are two potential directions to accelerate ML training speed. One is implementing ML algorithms in hardware, which should result in low power consumption and high efficiency. The other option is to reduce the complexity of ML algorithms while keeping a reasonable accuracy. 

\subsection{Light-weight ML Algorithms for Universal Embedded Systems}
Existing ML algorithms mainly focus on computer vision, natural language processing, and robotics with powerful graphics processing unit (GPU) or central processing unit (CPU) enabled computing to operate in real. However, communication systems are full of resource-constrained devices, e.g., embedded and Internet of things (IoT) systems. Therefore, the ML algorithms for communications should not only learn complex statistical models that underlie networks, consumers, and devices, but also effectively work with embedded devices having limited storage capabilities, computational power, and energy resources. It is challenging yet highly rewarding to develop light-weight ML algorithms, especially deep learning models, for embedded systems.

In this aspect, one potential direction will be the combination of ML and distributed computing frameworks such as fog computing and edge computing. Another important research direction is the investigation of high-level ML development library and toolbox.

\section{AI/ML for B5G Networks in Standards and study groups}\label{Sec6}
Although the convergence of AI/ML and communication networks is rapidly progressing, it is still in the early stage. As the various sensors, devices, applications, and systems connected in B5G networks will produce a variety of formats and sizes of data to be transmitted, it is extremely complex to standardize ML algorithms for B5G networks. No standard or baseline ML algorithm has been established and it is unclear for the whole communication community which types of ML algorithms suit the B5G systems best. 

Recently, there have been some preliminary works on applying AL/ML to B5G networks in standards including ITU and 3GPP, as well as other study groups such as FuTURE, telecom infra project (TIP), and 5G PPP, as shown in Table \ref{tab:standards}. ITU started a focus group on ``Machine learning for future networks including 5G (ML5G)"\footnote{\url{https://www.itu.int/en/ITU-T/focusgroups/ml5g/Pages/default.aspx} [Accessed: 26-June-2018]} by ITU-T Study Group 13 at its meeting in Geneva, 6--17 November 2017. The focus group will draft technical reports and specifications for ML for future networks, including interfaces, network architectures, protocols, algorithms, and data formats. The three working groups are ``Use cases, services and requirements", ``Data formats \& ML technologies", and ``ML-aware network architecture". 3GPP standards group developed a ML function that could allow 5G operators to monitor the status of a network slice or third-party application performance on ``Zero Touch \& Carrier Automation Congress"\footnote{\url{http://www.tech-invite.com/3m29/tinv-3gpp-29-520.html} [Accessed: 26-June-2018]} in Madrid, 22 March 2018. The network data analytics function (NWDAF) forms a part of the 3GPP's 5G standardization efforts and could become a central point for analytics in the 5G core network. Note that the NWDAF is still in the ``early stages" of standardization but could become ``an interesting place for innovation". A white paper named ``Wireless big data for smart 5G"\footnote{\url{http://www.future-forum.org/en/} [Accessed: 26-June-2018]} was published on FuTURE forum in November 2017. This white paper is a collection of pioneering research works on big data for 5G in China, both in academic and industry. It proposed the concept of ``smart 5G" and believed that 5G network needs to embrace new and cutting-edge technologies such as wireless big data and AI to efficiently boost both spectrum efficiency and energy efficiency, improve the user experience, and reduce the cost. TIP launched a project group ``AI and applied machine learning"\footnote{\url{http://telecominfraproject.com/introducing-the-tip-artificial-intelligence-and-applied-machine-learning-project-group/} [Accessed: 26-June-2018]} in November 2017. It will apply AI and ML to network planning, operations, and customer behavior identification to optimize service experience and increase automation. The objective is to define and share reusable, proven practices, models and technical requirements for applying AI and ML to reduce the cost of planning and operating telecommunications networks, understand and leverage customer behavior, and optimize service quality for an improved experience. 5G PPP also launched its efforts on combining AI with wireless communications, such as CogNet  \footnote{\url{http://www.cognet.5g-ppp.eu/cognet-in-5gpp/} [Accessed: 16-September-2019]}. It aims to build an intelligent system of insights and action for 5G network management. These developments in standards and study groups aim to use AI for physical layer and network management, which will greatly boost the performance of wireless networks.

\begin{table}[tb!]
\caption{A summary of AI/ML for B5G networks in standards.}
\label{tab:standards}
\centering
  \arraybackslash
\begin{tabular}{|m{1.2cm}<{\centering}|m{1.2cm}<{\centering}|m{1.2cm}<{\centering}|m{4.7cm}<{\centering}|m{5.7cm}<{\centering}|}
\hline
\textbf{Leading organization}&\textbf{Group name}&\textbf{Starting time}&\textbf{Purpose}&\textbf{Applications}\\
\hline
3GPP&NWDAF&March 2018&Allow 5G operators to monitor the status of a network slice or third-party application performance&5G core network data analytics \\
\hline
5G PPP&CogNet&July 2015&Build an intelligent system of insights and action for 5G network management&Autonomic network management based on machine learning\\
\hline
FuTURE&Wireless big data for smart 5G&November 2017&A white paper that collects pioneering research works on big data for 5G in China&Boost spectrum efficiency and energy efficiency, improve the user experience, and reduce the cost\\
\hline
ITU&ML5G&November 2017&Identify relevant gaps and issues in standardization activities related to ML for future networks&Interfaces, network architectures, protocols, algorithms, and data formats\\
\hline
TIP&AI and applied machine learning&November 2017&Define and share reusable, proven practices, models, and technical requirements for applying AI and ML&ML-based network operations, optimization, and planning; Customer behavior-driven service optimization; Multi-vendor ML-AI data exchange formats\\
\hline
\end{tabular}
\end{table}

\section{Conclusions and future challenges}\label{Sec7}
In this article, we have investigated how AI and ML can be used to efficiently solve the unstructured and seemingly intractable problems in future B5G wireless communication networks. A comprehensive survey of recent advances on the combination of AI/ML and wireless networks has been provided, including channel measurements, modeling, and estimation, physical-layer research, and network management and optimization. Challenges and potential future research directions have been discussed. AI algorithms and their applications to B5G networks have been introduced. An overview of developments for applying AI/ML to B5G systems carried out by standard organizations and study groups have also been provided.

\clearpage
\section*{Biographies}
\textbf{Cheng-Xiang Wang} [S'01-M'05-SM'08-F'17] (chxwang@seu.edu.cn) received his Ph.D. degree from Aalborg University, Denmark, in 2004. He has been with Heriot-Watt University, Edinburgh, U.K., since 2005, where he was promoted to a Professor in 2011. In 2018, he joined Southeast University, China, as a professor. He is also a part-time professor with the Purple Mountain Laboratories, Nanjing, China. He has authored three books, one book chapter, and more than 360 papers in refereed journals and conference proceedings, including 23 Highly Cited Papers. His current research interests include wireless channel measurements and modeling, B5G wireless communication networks, and applying artificial intelligence to wireless communication networks. He is a Fellow of the IET, an IEEE Communications Society Distinguished Lecturer, in 2019 and 2020, and a Highly-Cited Researcher recognized by Clarivate Analytics, in 2017-2019. He is currently an Executive Editorial Committee Member of the IEEE TRANSACTIONS ON WIRELESS COMMUNICATIONS. He received ten Best Paper Awards from IEEE GLOBECOM 2010, IEEE ICCT 2011, ITST 2012, IEEE VTC 2013-Spring, IWCMC 2015, IWCMC 2016, IEEE/CIC ICCC 2016, WPMC 2016, and WOCC 2019.

\textbf{Marco Di Renzo} [F'20] (marco.direnzo@centralesupelec.fr) received the Ph.D. degree in electrical engineering from University of L'Aquila, Italy, in 2007, and the D.Sc. degree (HDR) from University Paris-Sud, France, in 2013. He is a CNRS Research Director (CNRS Professor), and a faculty member of  CentraleSupelec and Paris-Saclay University, Paris, France. Also, he is a Nokia Foundation Visiting Professor at Aalto University, Helsinki, Finland. He is a Highly Cited Researcher, an IEEE Fellow, and a Distinguished Lecturer of the IEEE Communications Society (COMSOC) and IEEE Vehicular Technology Society. Currently, he serves as the Editor-in-Chief of IEEE Communications Letters. He has received several awards, including the SEE-IEEE Alain Glavieux Award, the IEEE Jack Neubauer Memorial Best Systems Paper Award, the IEEE COMSOC Young Professional in Academia Award, the IEEE COMSOC Best Young Researcher Award, and the Best Paper Award at IEEE ICC 2019. 

\textbf{Slawomir Sta\'nczak} [M'04-SM'11] (slawomir.stanczak@hhi.fraunhofer.de) studied electrical engineering with specialization in control theory at the Wroclaw University of Technology and at the Technical University of Berlin (TU Berlin). He received the Dipl.-Ing. degree in 1998 and the Dr.-Ing. degree (summa cum laude) in electrical engineering in 2003, both from TU Berlin; the Habilitation degree (venialegendi) followed in 2006. Since 2015, he has been a Full Professor for network information theory with TU Berlin and the head of the Wireless Communications and Networks department at Fraunhofer Institute for Telecommunications, Heinrich Hertz Institute (HHI). Prof. Stanczak  is a co-author of two books and more than 200 peer-reviewed journal articles and conference papers in the area of information theory, wireless communications, signal processing and machine learning.He was an Associate Editor of the IEEE Transactions on Signal Processing between 2012 and 2015. Since February 2018 Prof. Stanczak has been the chairman of the ITU-T focus group on machine learning for future networks including 5G.   

\textbf{Sen Wang} (s.wang@hw.ac.uk) is an Assistant Professor in Robotics and Autonomous Systems at Heriot-Watt University and a faculty member of the Edinburgh Centre for Robotics. He is also the director of the Perception and Robotics Group in the Ocean Systems Laboratory. Previously, he was a post-doctoral researcher at the University of Oxford. His research interests include robotics, computer vision, and machine/deep learning.

\textbf{Erik G. Larsson}  (erik.g.larsson@liu.se) is Professor at Link\"oping University, Sweden, 
and a Fellow of the IEEE.     He   co-authored   \emph{Fundamentals of Massive MIMO} (Cambridge, 2016) and \emph{Space-Time 
Block Coding for Wireless Communications} (Cambridge, 2003). 
Recent service includes membership of  the IEEE Signal Processing Society  
Awards Board (2017--2019),    the  \emph{IEEE Signal Processing Magazine} editorial board  (2018--2020), and the \emph{IEEE Transactions on Wireless Communications} steering committee 
(2019--2022). 
He received the  {IEEE Signal Processing Magazine} Best 
Column Award twice, in 2012 and 2014, the IEEE ComSoc Stephen 
O. Rice Prize in Communications Theory  2015, 
 the IEEE ComSoc Leonard G. Abraham 
Prize  2017, the IEEE ComSoc Best Tutorial Paper Award 2018 and 
the IEEE ComSoc Fred W. Ellersick Prize in 2019.

\newpage

\clearpage

\clearpage
\clearpage
\clearpage

\end{document}